\documentclass[prd,nofootinbib,twocolumn]{revtex4}
\usepackage{graphicx}
\usepackage{bm}
\usepackage{float}
\usepackage{subfigure}
\usepackage{amsmath}
\usepackage{amsfonts}
\usepackage{color}
\usepackage{slashed}
\usepackage{fancyhdr}
\def\be{\begin{equation}}
\def\ee{\end{equation}}
\def\bed{\begin{description}}
\def\eed{\end{description}}

\def\bea{\begin{eqnarray}}
\def\eea{\end{eqnarray}}

\def\ba{\begin{array}}
\def\ea{\end{array}}

\def\u1{$U(1)$}
\def\suu1{$SU(2)\times U(1)$}

\newcommand{\dx}[1]{\text{d}#1}

\newcommand{\dd}{\text{d}}
\newcommand{\mpl}{M_\text{pl}}
\newcommand{\GN}{G_\text{N}}

\begin{document}

\title{Gravitational Scattering Of Photons Off Cosmic Strings}

\author{Yi-Zen Chu$^1$ and Tanmay Vachaspati$^2$}
\affiliation{
$^1$Center for Particle Cosmology, Department of Physics and Astronomy,
University of Pennsylvania, Philadelphia, PA 19104, \\
$^2$Physics Department, Arizona State University,
Tempe, AZ 85287, USA.}

\begin{abstract}
\noindent
Photons can gravitationally scatter off a cosmic string loop and gain or lose energy. We consider the spectral distortion induced by cosmic string loops placed in an ambient thermal bath of photons. The fractional deviation from a thermal spectrum caused by cosmic strings is estimated to scale as $(\GN \mu)^2 z^2$, where $\GN$ is Newton's constant, $\mu$ is the string tension, and $z$ is the cosmological redshift after which spectral distortions can survive. This effect is large enough to potentially be of observational interest.
\end{abstract}

\maketitle

\section{Introduction}

The spectrum of cosmic microwave background (CMB) photons has been measured
very accurately and, at the present time, shows no deviations from a blackbody 
distribution. These measurements provide tight constraints on any source of energy
injection into the primordial medium at redshifts $z \lesssim 10^6$. 

The relativistic motion of cosmic strings may continually inject energy into the cosmic
medium starting from very early epochs, and certainly much earlier than $z=10^6$.
Depending on the particle physics interactions between the string and the cosmic
medium, the energy deposited could be significant 
\cite{Tashiro:2012nb}. However, there is also energy deposition due to the gravitational
interactions of photons and strings, and this process is completely generic. The
gravitational scattering of photons by strings can result in gravitational lensing of
background structure with angular separation $\sim 4\pi \GN\mu$ where $\GN$
is Newton's constant and $\mu$ is the string's lineal energy density.  However,
the bending of light is an elastic process and cannot lead to spectral
distortions. Photon energies can only be re-distributed by cosmic strings at
order $(\GN\mu)^2$ and this is the effect we are interested in analyzing. 

Inelastic scattering by strings is similar to photon pair creation by the time-dependent 
metric of a cosmic string as discussed in \cite{Steer:2010jk}. Indeed the
Feynman diagram is the same for both processes except for the reversal of the
momentum in an external photon leg. However, important technical differences
arise due to the different kinematics that make the scattering calculation more
difficult. To obtain the spectral distortions we also need to evaluate a 
Boltzmann integral which makes the analysis challenging.

In Sec.~\ref{Setup} we introduce the formalism and evaluate the scattering
amplitude for a general cosmic string loop. To obtain numerical results, we
specialize to a simple loop trajectory, one that is expected
to be close to the most generic loop \cite{Copi:2010jw}, and is
called a ``degenerate, kinky loop''.  In Sec.~\ref{KinkyLoop} we numerically
evaluate the photon scattering rate in phase space after averaging over all 
possible orientations and shapes of the degenerate kinky loop. The spectral distortion is
shown in Fig.~\ref{BoltzmannIntegral_Plot} and the result is used to estimate the
fractional spectral distortions of the CMB in Sec.~\ref{conclusions}. In 
Appendix~\ref{NIntegrateDiscussion} we discuss some technical details 
of the evaluation of the Boltzmann integrals.

\section{Setup}
\label{Setup}

We will consider the process of photons scattering off a cosmic string taking place in a weakly curved spacetime, 
\begin{align}
g_{\mu\nu} &= \eta_{\mu\nu} + \frac{h_{\mu\nu}}{\mpl}, \qquad \mpl \equiv 1/\sqrt{32\pi \GN}, \\
\eta_{\mu\nu} &= \text{diag}[1,-1,-1,-1],
\end{align}
The scattering process is governed by the action
\begin{align}
\label{Action}
S &\equiv S_\gamma + S_h + S_\text{NG} \\
&
- \frac{1}{2\mpl} \left( 
\int h_{\mu\nu} T^{\mu\nu}_{(\gamma)} \dd^4 x 
+ \int_0^L \dd\sigma \int \dd t \ h_{\mu\nu} T^{\mu\nu}_\text{(NG)}
\right),\nonumber
\end{align}
where $S_\gamma$ is the action of the photon $A_\mu$ 
\begin{align}
\label{Action_Photon}
S_\gamma &= -\frac{1}{4} \int F_{\mu\nu} F^{\mu\nu} \dd^4 x ,
\end{align}
with $F_{\mu\nu} \equiv \partial_\mu A_\nu - \partial_\nu A_\mu $; $S_h$ is the quadratic action of the graviton $h_{\mu\nu}$ in de Donder gauge $(\partial^\mu h_{\mu\nu} = (1/2) \partial_\nu h)$, 
\begin{align}
\label{Action_Graviton}
S_h &= \frac{1}{2} \int \dd^4 x \left( \partial^\mu h^{\alpha\beta} \partial_\mu h_{\alpha\beta} - \frac{1}{2} \partial^\mu h \partial_\mu h \right)
\end{align}
with $h \equiv \eta^{\mu\nu} h_{\mu\nu}$, and $S_\text{NG}$ is the Nambu-Goto action for the cosmic string
\begin{align}
\label{Action_String}
S_\text{NG} = -\mu \int_0^L \dd\sigma \int \dd t \sqrt{-\gamma}.
\end{align}
Here, our cosmic string is of length $L$ with tension $\mu$, and $X^\mu(t,\sigma)$ is the 4 coordinate vector of a given point $(t,\sigma)$ on its worldsheet. The determinant of the induced metric of the latter is $\gamma$; denoting 
\begin{align}
\dot{X}^\mu \equiv \partial_t X^\mu, \qquad (X^\mu)' \equiv \partial_\sigma X^\mu,
\end{align}
we have
\begin{align}
\gamma &= \dot{X}^2 X'^2 - \left(\dot{X}\cdot X'\right)^2 .
\end{align}
The stress-energies of the photons and the string are, respectively,
\begin{align}
\label{StressEnergy_Photons}
T^{\mu\nu}_{(\gamma)}
&= -F^{\mu\alpha} F^\nu_{\phantom{\mu}\alpha} + \frac{1}{4} \eta^{\mu\nu} F_{\alpha\beta} F^{\alpha\beta},
\end{align}
and
\begin{align}
\label{StressEnergy_String}
T^{\mu\nu}_\text{(NG)}
&= \mu \left( \dot{X}^\mu \dot{X}^\nu - (X^\mu)' (X^\nu)' \right).
\end{align}
For $T^{\mu\nu}_\text{(NG)}$, we have already chosen set of coordinates such that the induced metric on the string is conformally flat and $\dot{X} \cdot X' = 0$. In the same gauge, the solution to the Nambu-Goto equations of motion from eq. \eqref{Action_String}, $(\partial_t^2 - \partial_\sigma^2) X^\mu = 0$, can always be expressed as a superposition of null left- and right-movers,
\begin{align}
X^\mu(t,\sigma) = \frac{1}{2}\left( X^\mu_+[\sigma_+] + X^\mu_-[\sigma_-] \right)
\end{align}
and
\begin{align}
X_\pm^\mu = \left( \sigma_\pm, {\bf X}_\pm(\sigma_\pm) \right), \qquad X'^2_\pm = 0;
\end{align}
where from henceforth we shall switch to using the variables
\begin{align}
\sigma_\pm \equiv t \pm \sigma
\end{align}
and their associated derivatives
\begin{align}
\partial_\pm \equiv \partial/\partial \sigma_\pm,
\end{align}
instead of $(t,\sigma)$. Moreover it is advantageous to exploit the periodicity of the trajectory, i.e. $X^\mu[t,\sigma \pm L] = X^\mu[t,\sigma]$, to Fourier decompose the stress energy of the cosmic string, eq. \eqref{StressEnergy_String}, into the form
\begin{align}
\label{StressEnergy_NG}
-\frac{i}{2\mpl} \widetilde{T}^{\mu\nu}_\text{(NG)}[k] &=
-\frac{i\mu L}{4 \mpl}
\sum_{\ell = -\infty}^\infty C_{(+|\ell)}^{\{\mu} C_{(-|\ell)}^{\nu\}} \nonumber\\
&\qquad \qquad
\times (2\pi) \delta\left( k_0 - \frac{4\pi\ell}{L} \right) .
\end{align}
(We denote $U^{\{\mu} V^{\nu\}} \equiv U^\mu V^\nu + U^\nu V^\mu$.) The $C^\alpha_{(\pm|\ell)}$s contain information about the trajectory of the cosmic string loop and are given by
\begin{align}
\label{Cplusminus}
C^\alpha_{(\pm|\ell)}(k) \equiv \int_0^L \frac{\dx{\sigma_\pm}}{L} \exp\left[ \frac{i}{2} k_\mu X_\pm^\mu[\sigma_\pm] \right] \partial_\pm X_\pm^\alpha ,
\end{align}
where
\begin{align}
\label{kDef}
k \equiv p'-s p ,
\end{align}
For the scattering problem at hand, $s=1$, so that the $C_\ell$s become their complex conjugate under the interchange of $p \leftrightarrow p'$: $C_{(\pm|\ell)}[p,p'] = \overline{C_{(\pm|\ell)}}[p',p]$. (The overbar denotes complex conjugation.) For the pair production process described in \cite{Steer:2010jk}, $s=-1$.

Let us denote the polarization vector of the incoming photon as $\epsilon_\mu[p]$ and that of the outgoing photon as $\epsilon'_\mu[p']$. In the Lorenz gauge $(\partial^\mu A_\mu = 0)$, which we shall use here, they satisfy $\epsilon \cdot p = \epsilon' \cdot p' = 0$. The stress-energy operator of the photons in eq. \eqref{StressEnergy_Photons}, Wick-contracted with the state $|p\rangle$ representing one incoming photon of momentum $p$ and with the state $\langle p' |$ representing one outgoing photon of momentum $p'$ is 
\begin{align}
\label{StressEnergy_Photon}
\widetilde{T}^{\alpha\beta}_\text{($\gamma$)}[p',p]
&= -\bigg(
p^{\{\alpha} p'^{\beta\}} \ \epsilon \cdot \bar{\epsilon}'
+ p \cdot p' \ \epsilon^{\{\alpha} \bar{\epsilon}'^{\beta\}} \\
&\qquad \qquad
- p^{\{\alpha} \bar{\epsilon}'^{\beta\}} \epsilon \cdot p'
- p'^{\{\beta} \epsilon^{\alpha\}} \bar{\epsilon}' \cdot p \nonumber\\
&\qquad \qquad
- \eta^{\alpha\beta} \left( p \cdot p' \ \epsilon \cdot \bar{\epsilon}' - p \cdot \bar{\epsilon}' \ p' \cdot \epsilon
\right) \bigg) . \nonumber
\end{align}

\begin{figure}
\begin{center}
\includegraphics[width=2in,angle=-90]{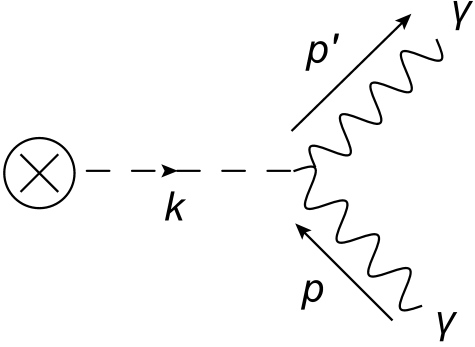}
\end{center}
\caption{Feynman diagram for the amplitude of photon scattering off a cosmic string in motion. 
The $\otimes$ represents the stress energy operator of a Nambu-Goto string in 
Eq.~\eqref{StressEnergy_String}. The dashed line is the graviton propagator in eq. \eqref{GravitonPropagator}. The rest of the diagram is the stress-energy of the photons in eq. \eqref{StressEnergy_Photon}; the wavy lines are the incoming photon with momentum $p$ 
and the outgoing photon with momentum $p'$. 
(Drawn with {\sf JaxoDraw} \cite{Binosi:2003yf}.)
}
\label{ScatteringDiagram}
\end{figure}

The propagator of the graviton that follow from eq. \eqref{Action_Graviton} is
\begin{align}
\label{GravitonPropagator}
D_{\alpha\beta \mu\nu}(k) 
= \frac{i}{2k^2}\left( \eta_{\alpha\mu} \eta_{\nu\beta} + \eta_{\alpha\nu} \eta_{\mu\beta} - \eta_{\mu\nu} \eta_{\alpha\beta} \right) .
\end{align}
At this point, the quantum amplitude of a single photon with initial momentum $p$, scattering off the cosmic string, and acquiring a final momentum $p'$ is then given by
\begin{align}
i\mathcal{M}(p \to p')
&= -8\pi G_\text{N} \widetilde{T}^{\mu\nu}_\text{(NG)} D_{\mu\nu\alpha\beta} \widetilde{T}^{\alpha\beta}_\text{($\gamma$)} .
\end{align}
(Because the classical stress-energy tensor of photons is traceless in 4 dimensions, one would find that $\widetilde{T}^{\mu\nu}_\text{(NG)} D_{\mu\nu\alpha\beta} \widetilde{T}^{\alpha\beta}_\text{($\gamma$)} = i \widetilde{T}^{\mu\nu}_\text{(NG)} \eta_{\mu\alpha} \eta_{\nu\beta} \widetilde{T}^{\alpha\beta}_\text{($\gamma$)}/k^2$.) This amplitude has a diagrammatic representation shown in 
Fig.~\ref{ScatteringDiagram}.

We proceed to calculate $|\mathcal{M}|^2$, summed over all initial and final photon polarizations. As a consequence of Ward identities obeyed by electromagnetism, this amounts to the replacements $\epsilon'_\mu \overline{\epsilon}'_\nu, \epsilon_\mu \overline{\epsilon}_\nu \to -\eta_{\mu\nu}$. From eq. \eqref{StressEnergy_NG}, we will also face the product $(2\pi)^2 \delta(k_0 - (4\pi\ell/L)) \delta(k_0 - (4\pi\ell'/L))$ and their associated sums over all integers $\ell$ and $\ell'$. This product is zero unless the arguments of the two $\delta$-functions are simultaneously equal to zero; this collapses the double sum into one, while the product of $\delta$ functions become 
\begin{align}
&\left( (2\pi) \delta(k_0 - (4\pi\ell/L)) \right)^2 \nonumber\\
&\to (2\pi) \delta(k_0 - (4\pi\ell/L))\text{(Total time elapsed)} .
\end{align}
Define the amplitude-squared per unit time as
\begin{align}
\left\vert\mathcal{A}(p \to p')\right\vert^2 \equiv \sum_\text{polarizations} \frac{\left\vert \mathcal{M}(p \to p') \right\vert^2}{2\pi \delta(0)},
\end{align}
where
\begin{align}
2\pi \delta(0) \equiv \lim_{\tau \to \infty} \int_{-\tau/2}^{+\tau/2} \dd t .
\end{align}
Suppressing the $\ell$ index on the $C_{(\pm|\ell)}$s, we arrive at
\begin{align}
\label{AmplitudeSquared_InfiniteSum}
\left\vert \mathcal{A} \right\vert^2
&= \sum_{\ell = -\infty}^\infty (2\pi) \delta\left( k_0 - \frac{4\pi\ell}{L} \right) \left\vert \mathcal{A}_\ell \right\vert^2 ,
\end{align}
with
{\allowdisplaybreaks\begin{widetext}
\begin{align}
\label{AmplitudeSquared_General}
\left\vert \mathcal{A}_\ell \right\vert^2
&\equiv \frac{(4 \pi G_\text{N} \mu L)^2}{(p' \cdot p)^2} 
\bigg(
8 |p' \cdot C_+|^2 |p' \cdot C_-|^2 \nonumber \\
&\qquad \qquad
- 4 s (p' \cdot p) \big\{
(p' \cdot C_+) (p' \cdot \overline{C_-}) (\overline{C_+} \cdot C_-) + (p' \cdot \overline{C_+}) (p' \cdot C_-) (C_+ \cdot \overline{C_-}) \nonumber \\
&\qquad \qquad \qquad \qquad
    - (p' \cdot \overline{C_+}) (p' \cdot \overline{C_-}) (C_+ \cdot C_-) - (p' \cdot C_+) (p' \cdot C_-) (\overline{C_+} \cdot \overline{C_-}) \nonumber \\
&\qquad \qquad \qquad \qquad
    + |C_-|^2 |p' \cdot C_+|^2 + |C_+|^2 |p' \cdot C_-|^2
\big\} \nonumber \\
&\qquad \qquad + 2 (p' \cdot p)^2 \big\{ |C_+ \cdot \overline{C_-}|^2 + |C_-|^2 |C_+|^2 - |C_+ \cdot C_-|^2 \big\}
\bigg),\qquad |C_\pm|^2 \equiv C^\alpha_\pm \overline{C^\beta_\pm} \eta_{\alpha\beta} 
\end{align}
\end{widetext}}
where $s$ is the same parameter occuring in \eqref{kDef}. We display this dependence of $|\mathcal{A}|^2$ on $s$ so as to compare against the pair production result in eq. (26) of \cite{Steer:2010jk}. The primary differences going from the scattering to pair production amplitude are that, $s$ changes sign from $1$ to $-1$, and the $k = p'-p$ in the $C_{(\pm|\ell)}$ is replaced with $k = p'+p$. For the pair production case, since $p_0, p'_0 \geq 0$, the sum over $\ell$ only runs over $0$ and the positive integers; whereas for the scattering case, $\ell$ runs over all integers and zero. Whichever $k=p'\pm p$ is being used, $k \cdot C_{(\ell|\pm)}(k) = 0$ always holds; we have used this to arrive at eq. \eqref{AmplitudeSquared_General}.

With the scattering amplitude per unit time, we may insert it into the generic form of the Boltzmann equation for the scattering of bosons (see, for example, Chapter 5 of \cite{KolbTurner}),
\begin{align}
\label{BoltzmannEquation_General}
&\frac{\dd^4 N_\gamma(t,{\bf p}')}{\dd t \dd^3 {\bf p}' (2\pi)^{-3}} 
= \frac{1}{2p'_0} \int\frac{\dd^3 {\bf p}}{(2\pi)^3 2p_0} \\
&\times\Big(
\left\vert\mathcal{A}({\bf p} \to {\bf p}')\right\vert^2 f(t,{\bf p})(1+f(t,{\bf p}')) \nonumber\\
&\qquad \qquad- \left\vert\mathcal{A}({\bf p}' \to {\bf p})\right\vert^2 f(t,{\bf p}')(1+f(t,{\bf p}))
\Big) \nonumber,
\end{align}
where $\dd^4 N_\gamma/\dd t \dd^3 {\bf p}' (2\pi)^{-3}$ is the number of photon scatterings off the cosmic string in question, per unit time, per unit momentum space; and $f(t,{\bf p})$ is the number of background photons per unit volume per unit momentum space. The first term describes the rate of photons entering the momentum bin ${\bf p}'$ due to their scattering from other momenta ${\bf p}$; the second term is the rate of leaving the momentum bin ${\bf p}'$ due to the same scattering mechanism. 

One may check directly using eq. \eqref{AmplitudeSquared_General} that $|\mathcal{A}_\ell|^2$ is invariant under the swap $p \leftrightarrow p'$: $p \cdot p'$ is invariant, whereas as noted already, $C_\pm[p',p] = \overline{C_\pm}[p,p']$, so $p' \cdot C_\pm$ is replaced with $p \cdot \overline{C_\pm} = -p' \cdot \overline{C_\pm}$ (and $p' \cdot \overline{C_\pm}$ with $-p' \cdot C_\pm$), where we have used $k \cdot C_\pm = 0$. Since every product of the $p' \cdot C_\pm$s and $p' \cdot \overline{C_\pm}$s contains even powers of $p'$, the minus sign can be dropped; hence, the swap $p \leftrightarrow p'$ amounts to the replacements $(p' \cdot C_\pm, p' \cdot \overline{C_\pm}) \to (p' \cdot \overline{C_\pm}, p' \cdot C_\pm)$ in every term of $|\mathcal{A}_\ell|^2$. The conclusion then follows if we observe that, the terms in $|\mathcal{A}_\ell|^2$ are either real or comes in pairs whose sum is real. This invariance $|\mathcal{A}_\ell({\bf p} \to {\bf p}')|^2 = |\mathcal{A}_\ell({\bf p}' \to {\bf p})|^2$ means the quadratic-in-$f$ terms in eq. \eqref{BoltzmannEquation_General} cancel. 

Furthermore, if we ignore the backreaction of this scattering process on the background distribution of photons, so that $f(t,{\bf p})$ can be assumed to be time independent; then $f(t,{\bf p}) = 2(\exp(|{\bf p}|/T_\gamma)-1)^{-1}$, 
the Bose-Einstein distribution for a massless spin-1 particle. For simplicity, let us also integrate over all possible final directions of the photon. Altogether, we are now lead to
\begin{align}
\label{BoltzmannEquation_PhotonsOffString}
&\frac{\dd^2 N_\gamma(t,|{\bf p}'|)}{\dd t \dd |{\bf p}'| (2\pi)^{-1}} \nonumber\\
&= \sum_{\ell=-\infty}^\infty \frac{|{\bf p}'|}{2(2\pi)^2}
\int_0^\infty \frac{\dd |{\bf p}| |{\bf p}|}{(2\pi)^3} \int_{\mathbb{S}^2} \dd\Omega_{\widehat{p}} \int_{\mathbb{S}^2} \dd\Omega_{\widehat{p}'} \\
&\times (2\pi)\delta\left( |{\bf p}'|-|{\bf p}| - \frac{4\pi\ell}{L} \right) \left\vert\mathcal{A}_\ell({\bf p} \to {\bf p}')\right\vert^2 \nonumber\\
&\times \left( \frac{1}{\exp(|{\bf p}|/T_\gamma)-1} - \frac{1}{\exp(|{\bf p}'|/T_\gamma)-1} \right) \nonumber
\end{align}
To be clear, our notation for the incoming $p$ and outgoing $p'$ momenta is
\begin{align}
p  = |{\bf p}| \left( 1, \widehat{p} \right), \qquad
p' = |{\bf p}'| \left( 1, \widehat{p}' \right)
\end{align}
so that, for instance $\dd \Omega_{\widehat{p}}$ is the integral over the solid angle of the incoming photon. 

Note that, because scattering does not change the total number of photons present, integrating the right hand side of eq. \eqref{BoltzmannEquation_PhotonsOffString} with respect to $|{\bf p}'|$ (from $0$ to $\infty$) would yield zero. We will later use this as a check of our numerical calculations.

To summarize what we have found so far: for a given cosmic string loop trajectory (and the relevant left- and right-movers $X_\pm$), first determine the modes $C_{(\pm|\ell)}$s in eq. \eqref{Cplusminus}, and afterwards insert them into eq. \eqref{AmplitudeSquared_General} to obtain $|\mathcal{A}_\ell|^2$. Then the rate of photon scattering off the cosmic string per unit energy, is obtained by evaluating the Boltzmann integral in eq. \eqref{BoltzmannEquation_PhotonsOffString}.

\section{Degenerate Kinky Loop}
\label{KinkyLoop}

In \cite{Steer:2010jk}, an analysis was performed for the gravitationally induced pair production rate of photons without specializing to a particular cosmic string loop trajectory. This is possible because, for the pair production case, there exists asymptotic expansions of eq. \eqref{Cplusminus} by utilizing the large dimensionless phase $|k \cdot X_\pm| \sim \ell \gg 1$. (In other words, one can exploit the large ratio of the astrophysical length of the string to the microscopic wavelength of the photon.) For instance, one may seek a saddle point of the $C_\pm$, i.e. search for configurations $(p,p')$ where $(p+p')\cdot X'_\pm = 0$. Because $p,p'$ and $X'_\pm$ are null vectors, that means $p+p'$ must be spacelike or null for a saddle point to exist. That in turn implies ${\bf p}$ and ${\bf p}'$ are collinear, since we require $p \cdot p' = |{\bf p}| |{\bf p}'|(1 - \widehat{p}\cdot\widehat{p}') \leq 0$. A slight deviation (in, say, the direction vectors $\widehat{p}$ and $\widehat{p}'$) from this configuration would fail to satisfy the stationary phase condition and therefore lead to a subdominant pair production rate; this in turn yields estimates on how very narrow regions (whose widths scale as some inverse power of $\ell$) in angular space provide the dominant contribution to the $C_\pm$ integrals. In all, what \cite{Steer:2010jk} found was that the energy emitted due to the pair of photons was largely independent of the mode number $\ell$; in other words, equal power in photon pairs were emitted at each harmonic $|{\bf p}| + |{\bf p}'| = 4\pi\ell/L$.

For the scattering case at hand, such an asymptotic analysis appears to be more difficult. For example, the stationary phase condition now becomes $(p'-p)\cdot X'_\pm = 0$; but with the minus sign instead of a plus sign (as was the case for pair production), the requirement that $p'-p$ is spacelike or null yields no useful constraint on the configuration $(p,p')$. Another important difference is that, for the pair production process, because $|{\bf p}'|+|{\bf p}| = 4\pi\ell/L$, the rate at a given $\ell$, without performing the analogous infinite sum in eq. \eqref{BoltzmannEquation_PhotonsOffString}, is already a physical quantity of interest; it is the power generated at a given energy $4\pi\ell/L$. For the scattering process, on the other hand, we wish to ask how many net photons are being added or subtracted, at a given $|{\bf p}'|$, from the background thermal bath. This means, for a fixed $|{\bf p}'|$ we have to perform both the solid angle integrals and the infinite sum over $\ell$ in eq. \eqref{BoltzmannEquation_PhotonsOffString} in order to obtain a physical quantity. However, at least for the two photon temperatures we have considered, we shall find that the dominant contribution to the infinite sum come from the small $|\ell |$ terms, specifically $|\ell | = 1,2,3,\dots,\mathcal{O}(10-100)$.

So as to achieve a concrete estimate of scattering rates of photons off cosmic strings, we shall therefore, in this section, specialize to the ``degenerate'' kinky loop trajectory that was also used in \cite{JonesSmith:2009ti} and \cite{Chu:2010zzb}.

Specifically, the left- and right-movers of the kinky loop are
\begin{align}
\label{LR_kinkyloop}
{\bf X}_+(\sigma_+) &= \left\{ \begin{array}{ll} 
\sigma_+ {\bf A} & 0 \leq \sigma_+ \leq \frac{L}{2} \\
(L-\sigma_+) {\bf A} & \frac{L}{2} \leq \sigma_+ \leq L \\
\end{array} \right. \nonumber \\
{\bf X}_-(\sigma_-) &= \left\{ \begin{array}{ll} \sigma_- {\bf B} & 0 \leq \sigma_- \leq \frac{L}{2} \\
(L-\sigma_-) {\bf B} & \frac{L}{2} \leq \sigma_- \leq L \\
\end{array} \right.
\end{align}
where ${\bf A}$ and ${\bf B}$ are constant unit vectors. Now, denoting
\begin{align*}
k_\text{A} \equiv k_i {\bf A}^i, \qquad
k_\text{B} \equiv k_i {\bf B}^i,
\end{align*}
we have, via  eq. \eqref{Cplusminus},
\begin{align}
C_{(+|\ell)}^\alpha &= \left( k_\text{A}, -k_0 {\bf A} \right) \frac{8 e^{i(k_\text{A}L/8-\pi\ell /2 )}}{L(k_\text{A}^2 - k_0^2)}
    \sin\left( \frac{k_\text{A}}{8}L-\frac{\pi}{2}\ell \right) \nonumber \\
C_{(-|\ell)}^\alpha &= \left( -k_\text{B}, k_0 {\bf B} \right) \frac{8 e^{i(k_\text{B}L/8+\pi \ell/2)}}{L(k_\text{B}^2 - k_0^2)}
		\sin\left( \frac{k_\text{B}}{8}L+\frac{\pi}{2}\ell \right)
\end{align}
We first integrate over $|{\bf p}|$ in eq. \eqref{BoltzmannEquation_PhotonsOffString}, using the $\delta$-function to set $|{\bf p}| = |{\bf p}'| - 4\pi\ell/L$ in the integrand. Because $|{\bf p}| \geq 0$, we also need to introduce a step function $\Theta(|{\bf p}'| - 4\pi\ell/L)$ multiplying the result after integration. Next, for convenience, we set $L=1$ and rescale
\begin{align}
\label{Rescaling}
(|{\bf p}'|, |{\bf p}|, T_\gamma) \to 4\pi(|{\bf p}'|,|{\bf p}|,T),
\end{align}
followed by the definitions
\begin{align}
\label{Rescaling_Start}
\vec{\Delta} \equiv {\bf p}' - {\bf p} = {\bf p}' - (|{\bf p}'|-\ell) \widehat{p} .
\end{align}
With the rescaling in eq. \eqref{Rescaling}, the length $|\vec{\Delta}|$, as $\widehat{p} \cdot \widehat{p}'$ is varied from $+1$ to $-1$, ranges over
\begin{align}
| \ell | = \left\vert |{\bf p}'| - |{\bf p}| \right\vert \leq |\vec{\Delta}| \leq |{\bf p}'| + |{\bf p}| = 2|{\bf p}'|-\ell .
\end{align}
Now, eq. \eqref{BoltzmannEquation_PhotonsOffString} reads
\begin{widetext}
\begin{align}
\label{BoltzmannIntegral_KinkyLoop_I}
\frac{\dd^2 N_\gamma(t,|{\bf p}'|)}{\dd t \dd |{\bf p}'| (2\pi)^{-1}} 
&= (\GN\mu)^2 \sum_{\substack{-\infty \leq \ell \leq \infty \\ \ell \neq 0 \\ |{\bf p}| \equiv |{\bf p}'|-\ell}} \int_{\mathbb{S}^2} \dd\Omega_{\widehat{p}'} \int_{\mathbb{S}^2} \dd\Omega_{\widehat{p}} 
\frac{256 |{\bf p}'| |{\bf p}| \Theta(|{\bf p}|)}{\pi^4 \left( 1-\widehat{p}\cdot\widehat{p}' \right)^2} \left( \frac{1}{e^\frac{|{\bf p}|}{T}-1} - \frac{1}{e^\frac{|{\bf p}'|}{T}-1} \right) \\
&\times \frac{\sin ^2\left(\frac{1}{2} \pi \left( {\bf A}\cdot\vec{\Delta} + \ell \right)\right)}{\left(({\bf A}\cdot\vec{\Delta} )^2 - \ell^2\right)^2} {\bf A}^i {\bf A}^j \left( \Delta^i \Delta^j (1+\widehat{p}\cdot\widehat{p}')-2\ell\left(|{\bf p}'| \widehat{p}'^i \widehat{p}'^j - |{\bf p}| \widehat{p}^i \widehat{p}^j \right) + \delta^{ij} \ell^2 (1-\widehat{p}\cdot\widehat{p}')\right) \nonumber\\
&\times \frac{\sin ^2\left(\frac{1}{2} \pi \left( {\bf B}\cdot\vec{\Delta} - \ell \right)\right)}{\left(({\bf B}\cdot\vec{\Delta} )^2 - \ell^2\right)^2} {\bf B}^k {\bf B}^l
\left( \Delta^k \Delta^l (1+\widehat{p}\cdot\widehat{p}') - 2\ell\left(|{\bf p}'| \widehat{p}'^k \widehat{p}'^l - |{\bf p}| \widehat{p}^k \widehat{p}^l \right) + \delta^{kl} \ell^2 (1-\widehat{p}\cdot\widehat{p}') \right) . \nonumber
\end{align}
\end{widetext}
{\it Collinear limit} \qquad The reader may worry that the presence of $(1-\widehat{p}\cdot\widehat{p}')^2$ in the denominator of eq. \eqref{BoltzmannIntegral_KinkyLoop_I} indicates the differential scattering rate may be singular in the collinear limit, as $\widehat{p}\cdot\widehat{p}' \to 1$. This is not the case, but rather,
\begin{align}
\label{BoltzmannIntegral_KinkyLoop_Collinear}
&\lim_{\widehat{p} \to \widehat{p}'} \frac{\dd^2 N_\gamma(t,|{\bf p}'|)}{\dd t \dd |{\bf p}'| (2\pi)^{-1}  \dd\Omega_{\widehat{p}'}  \dd\Omega_{\widehat{p}}} \\
&= (\GN\mu)^2 \sum_\ell \frac{256 |{\bf p}'| |{\bf p}| \Theta(|{\bf p}|)}{\pi^4 \ell^4} \left( \frac{1}{e^\frac{|{\bf p}|}{T}-1} - \frac{1}{e^\frac{|{\bf p}'|}{T}-1} \right) \nonumber\\
&\qquad \qquad
\times 
\frac{\sin^2\left( \frac{\pi\ell}{2} \left(1-{\bf A}\cdot\widehat{p}'\right) \right)}{1-\left({\bf A}\cdot\widehat{p}'\right)^2}
\frac{\sin^2\left( \frac{\pi\ell}{2} \left(1-{\bf B}\cdot\widehat{p}'\right) \right)}{1-\left({\bf B}\cdot\widehat{p}'\right)^2} .
 \nonumber
\end{align}
The third line of eq. \eqref{BoltzmannIntegral_KinkyLoop_Collinear} contains all the remaining dependence on angles and is finite for all ${\bf A}$, ${\bf B}$, and $\widehat{p}'$, because $\ell$ is an integer and $\sin(x)/x$ is finite at $x=0$. 

{\it $\ell=0$ Can Be Dropped} \qquad We also note that the $\ell=0$ term in eq. \eqref{BoltzmannIntegral_KinkyLoop_I} can in fact be dropped because, in this limit, the Bose-Einstein factors on the first line of eq. \eqref{BoltzmannIntegral_KinkyLoop_I} goes to zero while the rest of the integrand is finite. Explicitly, the Bose-Einstein factors, when $|\ell |/T \ll 1$, can be expanded as 
\begin{align}
\label{BoseEinstein_Smallell}
&\lim_{\ell\to 0}\left( \frac{1}{e^\frac{|{\bf p}|}{T}-1} - \frac{1}{e^\frac{|{\bf p}'|}{T}-1} \right)
=  \frac{\ell}{T} \frac{e^{|{\bf p}'|/T}}{\left( e^{|{\bf p}'|/T} - 1 \right)^2} \\
&\times \Bigg( 1 + \frac{\ell}{T} \frac{\coth\left(\frac{|{\bf p}'|}{2 T}\right)}{2}
+ \left(\frac{\ell}{T}\right)^2 \frac{\left(3 \text{csch}^2\left(\frac{|{\bf p}'|}{2 T}\right)+2\right)}{12} \nonumber\\
&\qquad
+ \left(\frac{\ell}{T}\right)^3 \frac{\left(\cosh \left(\frac{|{\bf p}'|}{T}\right)+5\right) \coth \left(\frac{|{\bf p}'|}{2 T}\right) \text{csch}^2\left(\frac{|{\bf p}'|}{2 T}\right)}{48} \nonumber\\
&\qquad
+\mathcal{O}(\ell^4)  \Bigg) . \nonumber
\end{align}
(For later use, we have expanded up to $\mathcal{O}(\ell^4)$.) This leads us to the following expansion of the small $\ell$ terms of eq. \eqref{BoltzmannIntegral_KinkyLoop_I},
\begin{align}
\label{BoltzmannIntegral_KinkyLoop_ell=0}
&(\GN\mu)^2 \left( \frac{1+\widehat{p}\cdot\widehat{p}'}{1-\widehat{p}\cdot\widehat{p}'} \right)^2 
\frac{256 |{\bf p}'|^2}{\pi^4} \frac{\ell e^{|{\bf p}'|/T}}{T \left( e^{|{\bf p}'|/T} - 1 \right)^2} \left(\frac{\pi}{2}\right)^4  \nonumber\\
&\times
\left( \frac{\sin\left( \frac{\pi}{2} |{\bf p}'| {\bf A} \cdot \left(\widehat{p}'-\widehat{p}\right) \right)}{\frac{\pi}{2} |{\bf p}'| {\bf A} \cdot \left(\widehat{p}'-\widehat{p}\right)} 
\frac{\sin\left( \frac{\pi}{2} |{\bf p}'| {\bf B} \cdot \left(\widehat{p}'-\widehat{p}\right) \right)}{\frac{\pi}{2} |{\bf p}'| {\bf B} \cdot \left(\widehat{p}'-\widehat{p}\right)} \right)^2 \nonumber\\
&\qquad \qquad + \mathcal{O}(\ell^2) .
\end{align}
For $\widehat{p}\cdot\widehat{p}' \neq 1$, this expression is non-singular for all $|{\bf p}'|$, and therefore zero when the limit $\ell \to 0$ is taken. We also remark that, in the preceding discussion on the collinear limit, if we took $\widehat{p}\cdot\widehat{p}' \to 1$ followed by $\ell\to 0$, we would find the Boltzmann integrand in eq. \eqref{BoltzmannIntegral_KinkyLoop_Collinear} would read
\begin{align}
&\lim_{\ell \to 0} \Bigg\{ (\GN\mu)^2 \frac{256 |{\bf p}'|^2}{\pi^4 \ell^4} \frac{\ell e^\frac{|{\bf p}|}{T}}{T (e^\frac{|{\bf p}|}{T}-1)^2} \\
&\times 
\left( \frac{\pi\ell}{2} \right)^2 \frac{1-{\bf A}\cdot\widehat{p}'}{1+{\bf A}\cdot\widehat{p}'}
\left( \frac{\pi\ell}{2} \right)^2 \frac{1-{\bf B}\cdot\widehat{p}'}{1+{\bf B}\cdot\widehat{p}'} 
\left( 1 + \mathcal{O}(\ell) \right) \Bigg\} = 0
 \nonumber
\end{align}
In other words, the two limits $\widehat{p}\cdot\widehat{p}' \to 1$ and $\ell \to 0$ commute and yield a zero integrand.

{\it Average Over Loop Configurations} \qquad If we were able to integrate over the solid angles with respect to the incoming and outgoing photon directions ($\widehat{p}$ and $\widehat{p}'$), the resulting expression must only depend on ${\bf A}$ and ${\bf B}$ through the Euclidean dot product ${\bf A} \cdot {\bf B}$, since the scattering rate is a scalar and there are no other vectors in the current problem. The result would allow one to see how the scattering rate per unit energy $|{\bf p}'|$ depends on the kinky loop configuration. However, in this paper, we shall be content with obtaining an estimate of the rate of scattering averaged over all possible kinky loop configurations -- if we assume a uniform distribution, this means we shall now examine
\begin{align}
\label{BoltzmannIntegral_KinkyLoop_Averaging}
\left\langle \frac{\dd^2 N_\gamma(t,|{\bf p}'|)}{\dd t \dd |{\bf p}'| (2\pi)^{-1}} \right\rangle
&\equiv \int_{\mathbb{S}^2} \frac{\dd\Omega_{{\bf A}}}{4\pi} \int_{\mathbb{S}^2} \frac{\dd\Omega_{{\bf B}}}{4\pi} 
\frac{\dd^2 N_\gamma(t,|{\bf p}'|)}{\dd t \dd |{\bf p}'| (2\pi)^{-1}} .
\end{align}
Notice the Boltzmann integrand in eq. \eqref{BoltzmannIntegral_KinkyLoop_I} has factorized into three portions: the first line is independent of ${\bf A}$ and ${\bf B}$; the second depends solely on ${\bf A}$ and the third solely on ${\bf B}$. This means the averaging in eq. \eqref{BoltzmannIntegral_KinkyLoop_Averaging} also factorizes into two independent solid angle integrals, one with respect to ${\bf A}$ and the other with respect to ${\bf B}$. In fact, the primary integrals we need are
\begin{align}
\label{Iij}
I_A^{ij} 
&\equiv \int_{\mathbb{S}^2} \frac{\dd\Omega_{{\bf A}}}{4\pi} \frac{\sin ^2\left(\frac{1}{2} \pi \left( {\bf A}\cdot\vec{\Delta} + \ell \right)\right)}{\left(({\bf A}\cdot\vec{\Delta} )^2 - \ell^2\right)^2} {\bf A}^i {\bf A}^j \\
I_B^{ij} 
&\equiv \int_{\mathbb{S}^2} \frac{\dd\Omega_{{\bf B}}}{4\pi} \frac{\sin ^2\left(\frac{1}{2} \pi \left( {\bf B}\cdot\vec{\Delta} - \ell \right)\right)}{\left(({\bf B}\cdot\vec{\Delta} )^2 - \ell^2\right)^2} {\bf B}^i {\bf B}^j \nonumber
\end{align}
By a flip of parity, ${\bf A} \to -{\bf A}$ or ${\bf B} \to -{\bf B}$, we see that these two integrals are actually the same, $I_A^{ij} = I_B^{ij} \equiv I^{ij}$. Therefore let us focus on the first. Observe that the only vector in the integrand is $\vec{\Delta}$, and the integral transforms as a rank 2 tensor under O$(3)$ transformations: for $\Delta^i \to U^i_{\phantom{i}j} \Delta^j$, where $U \in$ O$(3)$, we have $I^{ij} \to U^i_{\phantom{i}j} U^k_{\phantom{k}l} I^{jl}$. Therefore we may fix the form of $I^{ij}$ to be
\begin{align}
I^{ij} = \Delta^i \Delta^j A_\Delta + \delta^{ij} A_\delta,
\end{align}
where $A_\Delta,~A_\delta$ are O$(3)$ scalars. Contracting both sides with $\delta_{ij}$ and with $\Delta_i \Delta_j$ allow us to solve for $A_\Delta$ and $A_\delta$ in terms of the two scalar integrals
\begin{align}
I_n &\equiv \int_{\mathbb{S}^2} \frac{\dd\Omega_{{\bf A}}}{4\pi} \frac{\sin ^2\left(\frac{1}{2} \pi \left( {\bf A}\cdot\vec{\Delta} + \ell \right)\right)}{\left(({\bf A}\cdot\vec{\Delta} )^2 - \ell^2\right)^n}, \ n = 1,2 \\
&= \frac{1}{2} \int_{-1}^{+1} \dd c \frac{\sin^2\left(\frac{1}{2} \pi \left( |\vec{\Delta}|c + \ell \right)\right)}{\left(\vec{\Delta}^2 c^2 - \ell^2\right)^n} . \nonumber
\end{align}
where $c \equiv {\bf A}\cdot \vec{\Delta}/|\vec{\Delta}|$. The results are:
\begin{align}
A_\Delta &= \frac{3 I_1 + 3 \ell^2 \ I_2 - \vec{\Delta}^2 \ I_2}{2 \vec{\Delta}^4} \\
A_\delta &= \frac{\vec{\Delta}^2 \ I_2 - I_1 - \ell^2 \ I_2}{2 \vec{\Delta}^2}
\end{align}
and
{\allowdisplaybreaks\begin{align}
\label{I_1_and_I_2}
2 I_1 &= \frac{1}{2 \ell |\vec{\Delta}|} \Bigg( \ln\left\vert \frac{\ell-|\vec{\Delta}|}{\ell+|\vec{\Delta}|} \right\vert \nonumber\\
&\qquad \qquad -\left(\text{Ci}(\pi|\ell-|\vec{\Delta}||) - \text{Ci}(\pi |\ell+|\vec{\Delta}||)\right) \Bigg) \nonumber\\
2 I_2 &= \frac{1}{4 \ell^3 |\vec{\Delta}|} \Bigg( \ln\left\vert \frac{\ell+|\vec{\Delta}|}{\ell-|\vec{\Delta}|} \right\vert \\
&\qquad \qquad -\left(\text{Ci}(\pi |\ell+|\vec{\Delta}||) - \text{Ci}(\pi |\ell-|\vec{\Delta}||)\right) \Bigg) \nonumber\\
&\qquad + \frac{\pi}{4 \ell^2 |\vec{\Delta}|} \left( \text{Si}(\pi(\ell+|\vec{\Delta}|)) - \text{Si}(\pi(\ell-|\vec{\Delta}|)) \right) \nonumber\\
&\qquad - \frac{\cos(\pi(|\vec{\Delta}|+\ell))-1}{2\ell^2(\ell^2-\vec{\Delta}^2)} . \nonumber
\end{align}}
The Si and Ci are the Sine and Cosine Integral respectively. Whereas the $\ln$ minus Ci terms may also be re-expressed in terms of Cin and the Euler-Mascheroni  constant $\gamma_\text{E}$ (see \cite{NIST}),
\begin{align}
\label{Cin}
\text{Cin}(z) = \ln(z) - \text{Ci}(z) + \gamma_\text{E},
\end{align}
which is an even power series -- this is why we may put an absolute value around the arguments of $\ln$ and Ci.

With the $I_{1,2}$ in eq. \eqref{I_1_and_I_2}, and recalling $|{\bf p}| \equiv |{\bf p}'|-\ell$, $\vec{\Delta} \equiv {\bf p}' - (|{\bf p}'|-\ell) \widehat{p}$, the Boltzmann integral in eq. \eqref{BoltzmannIntegral_KinkyLoop_I} averaged over loop configurations now becomes
\begin{widetext}
\begin{align}
\label{BoltzmannIntegral_KinkyLoop_II}
\left\langle\frac{\dd^2 N_\gamma(t,|{\bf p}'|)}{\dd t \dd |{\bf p}'| (2\pi)^{-1}} \right\rangle
&= (\GN\mu)^2 \sum_{\substack{\ell = -\infty \\ |{\bf p}| \equiv |{\bf p}'|-\ell}}^{+\infty} \int_{-1}^{+1} \dd c'
\frac{512 |{\bf p}'| |{\bf p}| \Theta(|{\bf p}|)}{\pi^2 \vec{\Delta}^4 \left( 1-\widehat{p}\cdot\widehat{p}' \right)^2} 
\left( \frac{1}{e^\frac{|{\bf p}|}{T}-1} - \frac{1}{e^\frac{|{\bf p}'|}{T}-1} \right) \\
&\times \Bigg(
\left( \sum_{i,j = 1}^3 \left( 3 I_1 + 3 \ell^2 \ I_2 - \vec{\Delta}^2 \ I_2 \right)\frac{\Delta^i \Delta^j}{\vec{\Delta}^2} + \left( \vec{\Delta}^2 \ I_2 - I_1 - \ell^2 \ I_2 \right) \delta^{ij} \right) \nonumber\\
&\qquad \qquad
\times \left( \Delta^i \Delta^j (1+\widehat{p}\cdot\widehat{p}')-2\ell\left(|{\bf p}'| \widehat{p}'^i \widehat{p}'^j - |{\bf p}| \widehat{p}^i \widehat{p}^j \right) + \delta^{ij} \ell^2 (1-\widehat{p}\cdot\widehat{p}')\right) 
\Bigg)^2 \nonumber
\end{align}
\end{widetext}
where $c' \equiv \widehat{p} \cdot \widehat{p}'$. Even though $\widehat{p}$ and $\widehat{p}'$ appear in the integrand, after all the Euclidean dot products are carried out, the integrand in eq. \eqref{BoltzmannIntegral_KinkyLoop_II} will only depend on them via the combination $\widehat{p}\cdot \widehat{p}'$. Again using O$(3)$ invariance, we have reduced the $\int_{\mathbb{S}^2} \dd\Omega_{\widehat{p}}$ integration to $2\pi$ times a one dimensional integral with respect to $c'$, after the replacement $\widehat{p}\cdot\widehat{p}' \to c'$. After this integral with respect to $c'$ there will not be any dependence on $\widehat{p}$ nor $\widehat{p}'$, and therefore we have replaced the solid angle integral with respect to $\widehat{p}'$ with $4\pi$.

{\it Zero $|{\bf p}'|$ limit} \qquad 
Taking the zero outgoing energy limit, $|{\bf p}'| \to 0$, of 
Eq.~(\ref{BoltzmannIntegral_KinkyLoop_II}) we get
\begin{align}
\label{BoltzmannIntegral_KinkyLoop_LowE}
\left\langle\frac{\dd^2 N_\gamma(t,|{\bf p}'|)}{\dd t \dd |{\bf p}'| (2\pi)^{-1}} \right\rangle 
&\to
-\frac{64 LT_\gamma (\GN\mu)^2}{\pi^3 } \sum_{\ell = 0}^\infty \frac{\text{Cin}^2(2\pi\ell)}{\ell^3} \nonumber\\
&\approx -220 (LT_\gamma) (\GN\mu)^2 
\end{align}
where $T_\gamma = 4\pi T$ in this formula denotes the original (un-rescaled) photon temperature
(see Eq.~(\ref{Rescaling})), and we have reinstated $L$. This non-zero scattering rate indicates that, with a sufficiently high cosmic string density, there could be a significant backreaction on the low energy photon distribution.

To get to eq. \eqref{BoltzmannIntegral_KinkyLoop_LowE} we have set $|\vec{\Delta}| \to |\ell|$, and 
$|{\bf p}| \to -\ell$. In particular, starting from eq. \eqref{I_1_and_I_2}, as $|{\bf p}'| \to 0$,
\begin{align}
I_1 &\to -\frac{\text{Cin}(2\pi \ell)}{4\ell^2} \\
I_2 &\to \frac{\text{Cin}(2\pi\ell) + \pi \ell \ \text{Si}(2\pi \ell)}{8\ell^4} .
\end{align}
As we shall soon see, the rate of scattering (away from $|{\bf p}'|=0$) will continue to be negative for low energy photons, before becoming positive at higher energies. The motion of cosmic strings kick photons from lower to higher energies. 

{\it Change-of-variables} \qquad 
We will now evaluate eq. \eqref{BoltzmannIntegral_KinkyLoop_II} numerically for $T = 10^{22}$ and $10^{26}$. 
For a photon temperature of $3$ K, this corresponds to loop sizes of $L(T = 10^{22}) = 3$ kpc and 
$L(T = 10^{26}) = 3 \times 10^4$ kpc respectively.
(For comparison, the diameter of the Milky Way is 30 kpc.)
Let us change our variables from $c'$ to $\Delta \equiv |\vec{\Delta}|$, i.e.
\begin{align}
c'			&= \frac{|{\bf p}'|^2 + (|{\bf p}'|-\ell)^2 - \Delta^2}{2|{\bf p}'|(|{\bf p}'|-\ell)} \\
\dd c' 	&= -\frac{\Delta \dd \Delta}{|{\bf p}'|(|{\bf p}'|-\ell)},
\end{align}
to obtain
\begin{widetext}
\begin{align}
\label{BoltzmannIntegral_KinkyLoop_III}
&\left\langle\frac{\dd^2 N_\gamma(t,|{\bf p}'|)}{\dd t \dd |{\bf p}'| (2\pi)^{-1}} \right\rangle
= (\GN\mu)^2 
\sum_{\substack{\ell = -\infty \\ \ell \neq 0}}^\infty \frac{8 \Theta(|{\bf p}'|-\ell)}{\pi^2}
\left( \frac{1}{e^\frac{|{\bf p}'|-\ell}{T}-1} - \frac{1}{e^\frac{|{\bf p}'|}{T}-1} \right) \int_{|\ell|}^{2|{\bf p}'|-\ell} \frac{\dd\Delta}{\Delta ^9} \\
&\times \Bigg(
\frac{1}{\ell} \left( \text{Cin}\left(\pi(\Delta-\ell)\right) - \text{Cin}\left(\pi(\Delta+\ell)\right) \right) \left(3 \Delta^4 - \Delta ^2 \left( \ell^2 + (2 |{\bf p}'| - \ell)^2 \right) + 3 \ell ^2 (2 |{\bf p}'| - \ell)^2\right) \nonumber\\
&\qquad \qquad - \left(3 (2 |{\bf p}'| - \ell)^2-\Delta ^2\right) \left(2 \Delta 
   \left((-1)^{\ell } \cos (\pi  \Delta )-1\right)+\pi  (\Delta^2-\ell^2) \left( \text{Si}\left( \pi(\Delta+\ell) \right) + \text{Si}\left( \pi(\Delta-\ell) \right) \right)\right)
   \Bigg)^2 \nonumber
\end{align}
\end{widetext}
Let us understand qualitatively the integrand in eq. \eqref{BoltzmannIntegral_KinkyLoop_III}, over the range of integration, $|\ell| \leq \Delta \leq 2|{\bf p}'|-\ell$. Cin$(\pi z)$ is an even function of $z$, starting from zero at $z=0$ and rising, for large $|z| \gg 1$, to go as $\ln(\pi z)$ (see eq. \eqref{Cin}). Consequently, the $\ell^{-1}\{ \text{Cin}\left(\pi(\Delta-\ell)\right) - \text{Cin}\left(\pi(\Delta+\ell)\right) \}$ in eq.\eqref{BoltzmannIntegral_KinkyLoop_III} is an even function of $\ell$, peaks at $\Delta = |\ell|$ and proceeds to fall off as $(2/\Delta)(\cos(\pi\Delta)-1) + \mathcal{O}((\ell/\Delta)^2)$ for $\Delta \gg \ell$. Hence, the second line of  eq. \eqref{BoltzmannIntegral_KinkyLoop_III} grows at most as a cubic polynomial in $\Delta$ for $\Delta \gg \ell$. Whereas Si$(\pi z)$ is an odd function that rises from zero at $z=0$ and asymptotes to $\pi/2$ for large $\pi z \gg 1$; and thus the $\text{Si}\left( \pi(\Delta+\ell) \right) + \text{Si}\left( \pi(\Delta-\ell) \right)$ term asymptotes to $\pi$ for $\Delta \gg \ell$. We see that the third line of eq. \eqref{BoltzmannIntegral_KinkyLoop_III} (inside the $(\dots)^2$), grows as a quadratic polynomial multiplied by $3 (2 |{\bf p}'| - \ell)^2-\Delta ^2$, which decreases as $\Delta \to 2 |{\bf p}'| - \ell$ from below. With the $1/\Delta^9$ (on the first line in eq. \eqref{BoltzmannIntegral_KinkyLoop_III}) multiplying the entire squared quantity, we therefore expect the integrand to peak close to the lower end of the integration $\Delta \sim |\ell|$ and fall off with increasing $\Delta$. Moreover, for fixed $|{\bf p}'|$, large $|\ell| \gg 1$ terms ought to be suppressed relative to their smaller $|\ell|$ counterparts. These observations will aid us in the ensuing numerical integration -- in Appendix~\ref{NIntegrateDiscussion} we discuss some of the technical issues we face. 

We remind the reader that the rescaling in eq. \eqref{Rescaling} is in force from equations \eqref{Rescaling_Start} through \eqref{BoltzmannIntegral_KinkyLoop_III}, except for eq. \eqref{BoltzmannIntegral_KinkyLoop_LowE}.

\begin{figure}
\includegraphics[width=3.5in]{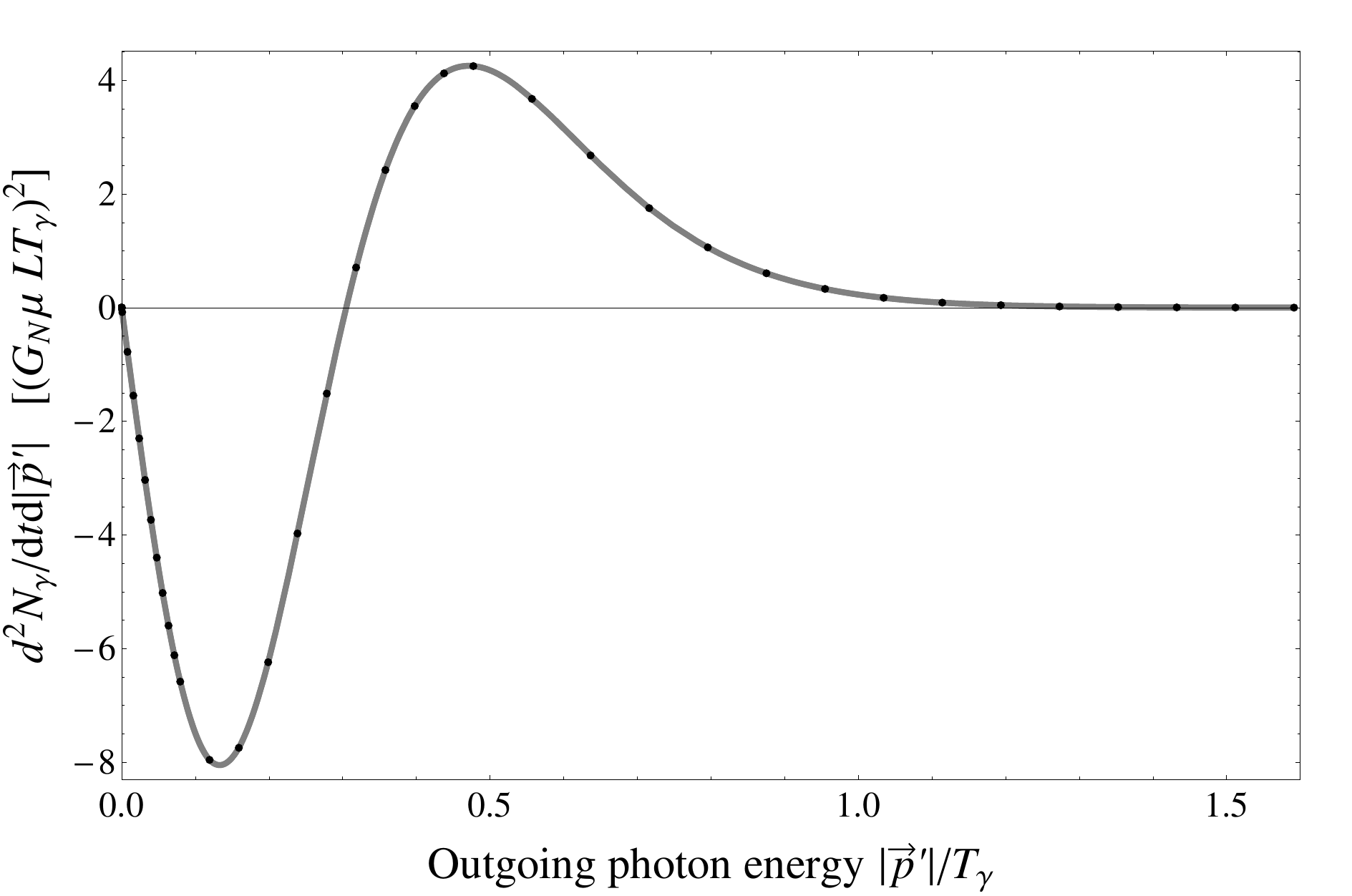}
\caption{The number of thermal photons scattering off the kinky cosmic string loop (of size $L$) described in eq. \eqref{LR_kinkyloop}, per unit time, per unit energy. There are actually two curves here, one at temperature $10^{22} (4\pi/L)$ and the other at $10^{26} (4\pi/L)$; because for $|{\bf p}'| \sim T_\gamma$ the rate scales as $(L T_\gamma)^2$, they coincide.}
\label{BoltzmannIntegral_Plot}
\end{figure}

In Fig.~\ref{BoltzmannIntegral_Plot} we present the results for eq. \eqref{BoltzmannIntegral_KinkyLoop_III}, when 
$T_\gamma = (4\pi/L) 10^{22}$ and $(4\pi/L) 10^{26}$, 
with respective maximum scattering rates per energy (roughly) $\pm 10^{47} (\GN \mu)^2$ and $\pm 10^{55} (\GN \mu)^2$. That the higher temperature yields a larger rate is what one would expect on physical grounds, since there are more photons available per unit volume at higher temperatures. (Photon number density is proportional to the cube of its temperature.) Since temperature enters the current problem only through the Bose-Einstein factors in eq. \eqref{BoltzmannEquation_PhotonsOffString}, and since it is these Bose-Einstein terms that are responsible for the existence of positive and negative scattering regions, we see that the energy scale over which this transition from negative to positive rate takes place is itself of the $\mathcal{O}(T_\gamma)$.

\section{Summary and Discussion}
\label{conclusions}

We have calculated the amplitude for inelastic scattering of photons off a cosmic string
loop. We have used the result to calculate the spectral distortions from a loop of string
placed in a thermal background of photons. The shape of the spectral distortion is shown
in Fig.~\ref{BoltzmannIntegral_Plot}.

It is of interest to estimate the fraction of photons that will be inelastically scattered by
a distribution of cosmic string loops. First we note that the scattering rate shown in
Fig.~\ref{BoltzmannIntegral_Plot}, can be estimated as
\be
\left\langle\frac{\dd^2 N_\gamma(t,\omega)}{\dd t \dd \omega} \right\rangle
\sim (\GN\mu)^2 (L T_\gamma)^2, \ \ \  \omega \sim T_\gamma,
\ee
where in this section we are using $\omega \equiv |{\bf p}'|$ to denote the frequency of the photon. Moreover, here we are only concerned with the scattering when $\omega \sim T_\gamma$ because this spectral region contains the most number of photons. The important scaling as $L^2$ has been verified numerically by comparing the two plots in Fig.~\ref{BoltzmannIntegral_Plot}; it can also be checked by using rescaling arguments on Eq.~\eqref{BoltzmannIntegral_KinkyLoop_I}.

Now we can multiply the scattering rate by the loop number density to get the number of scattered photons per unit time per unit volume. Since the scattering rate is proportional to $L^2$, the largest loops will be most important for this process. Such loops have size comparable to the horizon and so we take $L \sim t$, whereas the loop number density is approximately $1/t^3$, where $t$ is the cosmic time. Then we can integrate over time to get the total change in number density of photons $\Delta n$ per frequency interval, 
\begin{eqnarray}
\frac{\dd (\Delta n)}{\dd\omega} 
              &\sim& (\GN\mu)^2 \int_{t_\text{initial}}^{t_\text{now}} \dd t \frac{t^2 T_\gamma^2(t)}{t^3} \nonumber \\
              &\sim& (\GN\mu)^2 \int_{t_\text{initial}}^{t_\text{now}} \dd t \frac{T_\gamma^2(t)}{t} \nonumber \\
              &\sim& (\GN\mu)^2 T_i^2,
              \label{dNdo}
\end{eqnarray}
where $T_i$ is the initial temperature from which we start counting the number of
scattered photons. (We have used the fact that the photon temperature scales inversely as the cosmological scale factor, $T_\gamma \propto 1/a(t)$, assumed $a(t)$ goes as some power law, and the initial scale factor is much smaller than its present value $a(t_\text{initial}) \ll a(t_\text{now})$.) To get the fractional change in photon number density at some frequency, $(\dd(\Delta n)/\dd \omega)/(\dd n/\dd \omega) \sim \Delta n/n$ evaluated at the present epoch, due to scattering by strings, we divide Eq.~\eqref{dNdo} by the number density of ambient photons at the present time with frequency $\omega \sim T_0$, i.e. $\dd n/\dd \omega \sim T_0^2$.
Then
\be
\frac{\Delta n}{n} \sim (\GN\mu)^2 z^2
\ee
where $z$ is the redshift after which the scattered photons do not get
re-thermalized by Compton scattering. We will not go into a detailed calculation
here but simply mention that $z$ is definitely higher than the recombination
redshift ($\sim 10^3$) and possibly $\sim 10^4$ \cite{Sunyaev:1980vz}.

With $G_\text{N}\mu \sim 10^{-7}$ and using $z \approx 10^4$,
the spectral distortion from inelastic scattering of photons by the gravitational field of 
strings is at a level $\sim 10^{-6}$. The distortions are similar to the $y-$distortions
of the CMB in that the total number of photons is unchanged. However, the
shape is different and almost completely resides in the low frequency part
of the spectrum. 

As in the case of gravitational wave emission \cite{Damour:2000wa},
cosmic strings produce a burst signal in addition to a stochastic signal. It is possible
that cusps and kinks on strings can up-scatter ambient photons in a burst, and
that this may provide another observational signature. We leave the exploration
of this possible signature for future work.

\begin{acknowledgments}
We thank Levon Pogosian and Hiroyuki Tashiro for comments. YZC was supported by funds from the University of Pennsylvania; and by the DOE while at Arizona State University where this work was initiated. TV was supported by the DOE at ASU. Some of this work was done while the authors were at the Institute for Advanced Study, Princeton, which we thank for its hospitality. Much of the numerical and analytic work in this paper was done with {\sf Mathematica} \cite{mathematica}.
\end{acknowledgments}

\appendix

\section{Evaluating the Boltzmann Integral Numerically}
\label{NIntegrateDiscussion}

In this appendix we describe some of the technicalities involved in the numerical evaluation of the Boltzmann integral in eq. \eqref{BoltzmannIntegral_KinkyLoop_III}. To begin, let us re-express eq. \eqref{BoltzmannIntegral_KinkyLoop_III} in the following schematic form:
\begin{align}
\label{SchematicBoltzmann}
\left\langle\frac{\dd^2 N_\gamma(t,|{\bf p}'|)}{\dd t \dd |{\bf p}'| (2\pi)^{-1}} \right\rangle 
&= (\GN\mu)^2 
\sum_{\substack{\ell = -\infty \\ \ell \neq 0}}^\infty \frac{8 \Theta(|{\bf p}'|-\ell)}{\pi^2} \\
&\times \mathcal{E}_\ell \int_{|\ell|}^{2|{\bf p}'|-\ell} \mathcal{I}_\ell \dd\Delta , \nonumber
\end{align}
with
\begin{align}
\mathcal{E}_\ell &\equiv \frac{1}{e^\frac{|{\bf p}'|-\ell}{T}-1} - \frac{1}{e^\frac{|{\bf p}'|}{T}-1} .
\end{align}
One may attempt to use the {\sf NIntegrate} function in {\sf Mathematica} to calculate numerically the $\dd\Delta$-integral portion of eq. \eqref{SchematicBoltzmann}, for a given $|{\bf p}'|$, over as many $\ell$s as possible, before summing them up. One would find that the $|\ell| \sim \mathcal{O}(1-\text{few})$ terms are the dominant ones, but these same terms are such that, the pair $\pm |\ell|$ yields $\dd \Delta$ integrals of nearly equal magnitude but opposite signs. It is therefore prudent to combine the positive and negative $\ell$ terms, for $|\ell| \leq |{\bf p}'|$, before integrating, i.e.
\begin{align}
&\left\langle\frac{\dd^2 N_\gamma(t,|{\bf p}'|)}{\dd t \dd |{\bf p}'| (2\pi)^{-1}} \right\rangle \nonumber\\
&= \frac{8 (\GN\mu)^2}{\pi^2} \left\{ \sum_{\ell = 1}^{\lfloor |{\bf p}'| \rfloor} \left( \mathcal{B}_{(1|\ell)} + \mathcal{B}_{(2|\ell)} \right) 
+ \sum_{\ell = \lfloor |{\bf p}'| \rfloor + 1}^{+\infty} \mathcal{B}_{(3|\ell)} \right\} , \nonumber
\end{align}
where
{\allowdisplaybreaks\begin{align}
\mathcal{B}_{(1|\ell)} &\equiv \int_{\ell}^{2|{\bf p}'|-\ell} \left( \mathcal{E}_\ell \mathcal{I}_\ell  
+ \mathcal{E}_{-\ell} \mathcal{I}_{-\ell} \right) \dd\Delta \\
\mathcal{B}_{(2|\ell)} &\equiv \int_{2|{\bf p}'|-\ell}^{2|{\bf p}'|+\ell} \mathcal{E}_{-\ell} \mathcal{I}_{-\ell} \dd\Delta, \\
\mathcal{B}_{(3|\ell)} &\equiv \int_{\ell}^{2|{\bf p}'|+\ell} \mathcal{E}_{-\ell} \mathcal{I}_{-\ell} \dd\Delta .
\end{align}}
Notice we have also changed the summation variable such that now $\ell$ is summed over all non-zero positive integers. 

For $\mathcal{B}_{(1|\ell)}$, it is often useful to expand the Bose-Einstein factors using the small $\ell$ limit in eq. \eqref{BoseEinstein_Smallell} before applying {\sf NIntegrate}. Specifically, by decomposing the integrand $\mathcal{I}_\ell$ into $\ell$-even plus $\ell$-odd terms, and denoting $\mathcal{I}_\pm \equiv \mathcal{I}_\ell \pm \mathcal{I}_{-\ell}$, we have
\begin{align}
\label{CancellationSmallEll}
&\mathcal{B}_{(1|\ell)} \equiv \frac{\ell}{T} \frac{e^{|{\bf p}'|/T}}{\left( e^{|{\bf p}'|/T} - 1 \right)^2} \int_{\ell}^{2|{\bf p}'|-\ell} \dd\Delta \\
&\times \Bigg(
\mathcal{I}_-
+ \frac{\ell}{T} \frac{\coth\left(\frac{|{\bf p}'|}{2 T}\right)}{2} \mathcal{I}_+ \nonumber\\
&\quad + \left(\frac{\ell}{T}\right)^2 \frac{\left(3 \text{csch}^2\left(\frac{|{\bf p}'|}{2 T}\right)+2\right)}{12} \mathcal{I}_- \nonumber\\
&\quad + \left(\frac{\ell}{T}\right)^3 \frac{\left(\cosh \left(\frac{|{\bf p}'|}{T}\right)+5\right) \coth \left(\frac{|{\bf p}'|}{2 T}\right) \text{csch}^2\left(\frac{|{\bf p}'|}{2 T}\right)}{48} \mathcal{I}_+ \nonumber\\
&\quad + \mathcal{O}\left( \ell^4 \right) \Bigg) \nonumber
\end{align}
Of course, there are only two distinct integrals here, namely $\int_\ell^{2|{\bf p}'|-\ell} \mathcal{I}_\pm \dd\Delta$; the temperature dependent pieces of $\mathcal{B}_{(1|\ell)}$ do not take part in the integration. What we are emphasizing is the cancellations that occur between the $\ell$-odd and $\ell$-even portions of $\mathcal{I}_\ell$ when summing over the originally positive and negative $\ell$ terms, at each order in the $\ell/T$ expansion. For all the cases where we used this expansion in eq. \eqref{CancellationSmallEll}, we have found that the sum of the third and fourth lines is always much less than the second line -- usually much less than a percent -- and therefore may be neglected.

For each $|{\bf p}'| > 10$, we evaluate $\mathcal{B}_{(1,2|\ell)}$ for $\ell=1,2,3,\dots,10$ and sum them up. For $\ell > 10$, we perform the integrals for ten $\ell$s per decade, for example, $\ell=20,30,\dots,100$ and $\ell=2000,3000,\dots,10000$, up to a few orders of magnitude beyond that of $T$. (For small $|{\bf p}'| \ll T$, due to numerical underflow we do not integrate beyond $\ell \sim T$ but we simply stop at some $\ell$ that yields $\mathcal{B}_{(1,2,3|\ell)}$ that are many tens of orders of magnitude below $\mathcal{B}_{(1,2|\ell = 1)}$.) Then we use the {\sf Mathematica} function {\sf Interpolation} to compute a best fit line to the resulting numerical data and proceed to approximate the summation for $\ell > 10$ by an integral over the fit range, $\sum_{\ell = 11}^{\ell_\text{UV}} \to \int_{\ell = 10}^{\ell_\text{UV}} \dd \ell$. We find that, for each $|{\bf p}'|$, the sum over the first ten $\ell=1,2,\dots,10$ is usually responsible for roughly 91 percent of $\dd^2 N_\gamma(|{\bf p}'|)/\dd t \dd |{\bf p}'|$.

When numerically evaluating $\mathcal{B}_{(1,2,3|\ell)}$ we have found it useful to not do it over the entire range of $\Delta$, but rather start from integrating from the lower limit $\Delta_1$ to $\min[10 \Delta_1,\Delta_2]$, where $\Delta_2$ is the upper limit (for instance, $\Delta_1 = \ell$ and $\Delta_2 = 2|{\bf p}'| - \ell$ for $\mathcal{B}_{(1|\ell)}$), then from $\Delta_1$ to $\min[100 \Delta_1,\Delta_2]$, etc., until the error incurred is less than one percent. This algorithm is justified because, as already explained after eq. \eqref{BoltzmannIntegral_KinkyLoop_III}, we expect our integrand to always peak near $\Delta \sim |\ell|$. Practically speaking, this saves {\sf Mathematica} considerable amount of time and effort when sampling the integrand.

As a check of our calculation, we fit a line through the numerical data points of our plots in 
Fig.~\ref{BoltzmannIntegral_Plot} and find that the area above the $|{\bf p}'|$-axis is equal to the area below the $|{\bf p}'|$-axis at the sub-percent level. This is both consistent with our demand for percent-level accuracy in the numerical integrations of $\mathcal{B}_{(1,2,3|\ell)}$, as well as the physical requirement that scattering processes do not change the net photon number.


\begin{thebibliography}{99}

%\cite{Tashiro:2012nb}
\bibitem{Tashiro:2012nb} 
  H.~Tashiro, E.~Sabancilar and T.~Vachaspati,
  %``CMB Distortions from Superconducting Cosmic Strings,''
  Phys.\ Rev.\ D {\bf 85}, 103522 (2012)
  [arXiv:1202.2474 [astro-ph.CO]].
  %%CITATION = ARXIV:1202.2474;%%

\bibitem{Steer:2010jk} 
  D.~A.~Steer and T.~Vachaspati,
  %``Light from Cosmic Strings,''
  Phys.\ Rev.\ D {\bf 83}, 043528 (2011)
  [arXiv:1012.1998 [hep-th]].
  %%CITATION = ARXIV:1012.1998;%%
  
  %\cite{Copi:2010jw}
\bibitem{Copi:2010jw} 
  C.~J.~Copi and T.~Vachaspati,
  %``Shape of Cosmic String Loops,''
  Phys.\ Rev.\ D {\bf 83}, 023529 (2011)
  [arXiv:1010.4030 [hep-th]].
  %%CITATION = ARXIV:1010.4030;%%

\bibitem{JonesSmith:2009ti} 
  K.~Jones-Smith, H.~Mathur and T.~Vachaspati,
  %``Aharonov-Bohm Radiation,''
  Phys.\ Rev.\ D {\bf 81}, 043503 (2010)
  [arXiv:0911.0682 [hep-th]].
  %%CITATION = ARXIV:0911.0682;%%

\bibitem{Chu:2010zzb} 
  Y.~-Z.~Chu, H.~Mathur and T.~Vachaspati,
  %``Aharonov-Bohm Radiation of Fermions,''
  Phys.\ Rev.\ D {\bf 82}, 063515 (2010)
  [arXiv:1003.0674 [hep-th]].
  %%CITATION = ARXIV:1003.0674;%%

\bibitem{Binosi:2003yf} 
  D.~Binosi and L.~Theussl,
  %``JaxoDraw: A Graphical user interface for drawing Feynman diagrams,''  
  Comput.\ Phys.\ Commun.\  {\bf 161}, 76 (2004) [hep-ph/0309015].  %%CITATION = HEP-PH/0309015;%%

  D.~Binosi, J.~Collins, C.~Kaufhold and L.~Theussl,
  %``JaxoDraw: A Graphical user interface for drawing Feynman diagrams. Version 2.0 release notes,''
  Comput.\ Phys.\ Commun.\  {\bf 180}, 1709 (2009) [arXiv:0811.4113 [hep-ph]].  %%CITATION = ARXIV:0811.4113;%%

\bibitem{KolbTurner}
	E.~Kolb, M.~Turner, ``The Early Universe,'' Westview Press (1994)

\bibitem{NIST}
	http://dlmf.nist.gov/6.2,
	http://dlmf.nist.gov/6.6

%\cite{Sunyaev:1980vz}
\bibitem{Sunyaev:1980vz} 
  R.~A.~Sunyaev and Y.~.B.~Zeldovich,
  %``Microwave background radiation as a probe of the contemporary structure and history of the universe,''
  Ann.\ Rev.\ Astron.\ Astrophys.\  {\bf 18}, 537 (1980).
  %%CITATION = ARAAA,18,537;%%


%\cite{Damour:2000wa}
\bibitem{Damour:2000wa} 
  T.~Damour and A.~Vilenkin,
  %``Gravitational wave bursts from cosmic strings,''
  Phys.\ Rev.\ Lett.\  {\bf 85}, 3761 (2000)
  [gr-qc/0004075].
  %%CITATION = GR-QC/0004075;%%

\bibitem{mathematica}
	Wolfram Research, Inc., Mathematica, Champaign, IL (2011).

\end{thebibliography}
\end{document}